  \documentclass[prl,aps,twocolumn,showpacs]{revtex4}
\usepackage[dvips]{graphicx}
\usepackage{dcolumn}%
\usepackage{amsmath}%
\usepackage{amsfonts}%
\usepackage{amssymb}
\providecommand{\U}[1]{\protect\rule{.1in}{.1in}}
\newcommand{\be}{\begin{equation}}
\newcommand{\ee}{\end{equation}}
\newcommand{\ba}{\begin{array}}
\newcommand{\ea}{\end{array}}
\newcommand{\nn}{ \nonumber}
\newcommand{\ds}{\displaystyle}
\begin{document}

\title{Specific features of electric charge screening in few-layer graphene films}

\author{Natalya A. Zimbovskaya
}

\affiliation
{Department of Physics and Electronics, University of Puerto 
Rico-Humacao, CUH Station, Humacao, Puerto Rico 00791, USA; \\
Institute for Functional Nanomaterials, University of Puerto Rico, San Juan, Puerto Rico 00931, USA}

\begin{abstract}
We present a non-linear Thomas-Fermi theory which describes the electric charge screening in the system including two charged substrate layers separated by a few-layered graphene film. We show that by increasing the charge at the interfaces, the system could be turned from the weak screening regime where the whole film responds to the external charge, to the strong screening regime where the external charge is screened by a surface charge distribution confined to the bounding graphene layers. The transition from weak to strong screening is shown to turn on relatively quickly, and it happens when the applied external charge/external field reaches a certain crossover magnitude. The possibilities for experimental observation of the predicted crossover are discussed.
   \end{abstract}

\pacs{72.15.Gd,71.18.+y}

\date{\today}
\maketitle

 The past decade has seen a proliferation of interest in experimental and theoretical studies of graphene. The interest mainly stems from striking properties of graphene including tunable carriers type and density and exceptionally high carrier mobility. A very significant advantage of graphene is its inherent two-dimensionality which makes graphene-derived nanomaterials a promising family for applications in building nanoelectronic devices with planar device architectures \cite{1,2,3}. Such applications require fabrication of few-layer graphene films (FLG) supported by/placed in between insulating substrate layers. Correspondingly, the effects of the substrate on the electronic properties of FLG must be thoroughly analyzed. Some theoretical and experimental works concerning this issue already exist (see e.g. Refs. \cite{4,5,6,7,8}). However, further studies are necessary to get better quantitative understanding of these effects. 

In the present work, we contribute to these studies by theoretically analyzing the charge exchange and electrostatic potential spatial distribution in the system which consists of a FLG film sandwiched in between  two charged substrate layers. We assume that $"z" $ denotes the coordinate perpendicular to the graphene layers in the FLG, which occupies the space where $ 0 < z < D. $ The number of layers is supposed to be large enough to satisfy the condition $ d \ll D $ where $ d $ is the distance between adjacent layers in the film. At the interfaces between the FLG and the substrates $(z =0,D, $ respectively) the substrates are characterized with the areal charge  densities $ \sigma_{1s} $ at $ z = 0 $ and $ \sigma_{2s} $ at $ z = D. $ Also, we introduce charge carrier's densities $ \sigma_i $ corresponding to the graphene layers in  the FLG. The index $ "i" $ takes on values from $ 1 $ to $ N \ (N $ being the total number of the layers in the pack). For convenience, in further analysis we  assume that all above mentioned charge carrier densities may take on either positive or negative values depending on the nature of the charge carriers associated with a certain layer. We attribute positive values to the areal densities of holes and negative ones for those of electrons, respectively.

 Due to the specific form of the dispersion relation for the charge carriers in a single graphene sheet, their quantum-mechanical kinetic energy per unit area is proportional to the areal charge-carriers density in power $ 3/2.$ For a $i$-th graphene layer included  in the pack, the kinetic energy per unit area of substrate is given by the expression:
\be
K_i = \frac{2}{3}\sqrt \pi \hbar v_F |\sigma_i|^{3/2}   \label{1}
\ee
where $ v_F $ is the Fermi velocity of the charge carriers. We remark that this expression differs from the well known result applicable to a conventional two-dimensional conducting system where $ K \sim \sigma^2. $ The kinetic energy density for the whole FLG is obtained by summation over the layers. 

The charge distribution in the FLG is determined by competition between the kinetic energy of the charge carriers and their interactions with the self-consistent electrostatic potential. This may be described employing a nonlinear Thomas-Fermi theory for the 
charge carriers in the continuum limit which is appropriate when $ d \ll D $ so that the areal charge  carriers density rather smoothly changes on a scale of the interlayer spacing. At the same time, one must keep in mind that the Thomas-Fermi theory does not treat the effects of quantum coherence between the nearby graphene layers, and it applies in the absence of the interlayer tunneling. Therefore, the Thomas-Fermi approach may be used only assuming that the adjacent graphene  sheets are decoupled. This assumption is justified when the local Fermi energies are sufficiently far from the charge neutrality point and their shifts significantly exceed hopping amplitudes characterizing the interlayer tunneling \cite{8}. This requires the distances between the adjacent layers in the FLG to be large enough to prevent the tunneling of the charge carriers. Also, the Tomas-Fermi approach could be employed if the considered graphene pack belongs to the family of rotationaly faulted FLGs. In these systems crystalline lattices of adjacent graphene sheets are misoriented (twisted) at random angles with respect to each other, as established in several recent experiments \cite{9,10,11,12,13}. Twisted graphene multilayers exhibit only weak manifestations of interlyaer hybridization and interlayer tunneling of charge carriers. 
Previously, Thomas-Fermi models were successfully used to describe the electrostatic interactions in graphite intercalate compounds \cite{14,15} as well as to analyze the intrinsic screening in the graphene multylayers \cite{8}.

Within the Thomas-Fermi approach, the energy of the charge carriers in the FLG pack includes the kinetic term $K ,$ the term $ U_{int} $ describing electrostatic interactions between graphene layers and the term $ U_0 $ which originates from the interactions between the FLG and the charged substrates. Both $U_{int} $ and $ U_0 $ are energy densities calculated per unit area of the substrate, as well as the kinetic term $ K. $ In the continuum limit these terms have the form:
\be
K = \frac{2\sqrt \pi}{3} \hbar v_F \int_0^D \frac{dz}{d} |\sigma(z)|^{3/2} \equiv \gamma \int_0^D \frac{dz}{d} |\sigma(z)|^{3/2},   \label{2}
      \ee
\be
U_{int} = - \frac{e^2}{4\epsilon_0} \int_0^D \frac{dz}{d} \int_0^D \frac{dz'}{d} \sigma(z) \sigma(z') |z - z'| ,  \label{3} 
            \ee
\be 
U_0 = -\frac{ e^2}{2\epsilon_0} \sigma_0 \int_0^D \frac{dz}{d} \sigma(z) z  
\label{4}
                         \ee            
where $\epsilon_0 $ is permittivity of the free space,  and $ \sigma_0 = \sigma_{1s} - \sigma_{2s}. $ In further analysis we assume that the areal charge densities on the substrates have the same magnitude and opposite signs. Then the  electroneutrality of the system is provided by the conditions:
\be
\int_0^D \frac{dz}{d} \sigma(z) = 0,\qquad 
\sigma_{1s} = -\sigma_{2s} = \frac{1}{2}\sigma_0.  \label{5}
\ee

As it was demonstrated in the earlier work \cite{8}, the expression for the charge-carriers density $ \sigma (z) $ could be derived by minimizing the grand thermodynamic potential for the FLG. Introducing the chemical potential of the system $ \mu $ and combining the expressions for the kinetic and potential energy given by Eqs. (\ref{2})-(\ref{4}), we can write out the following expression for the density of the grand potential $ \Omega: $ 
\begin{align}
\Omega =& \int_0^D \frac{dz}{d} \bigg\{\gamma |\sigma(z)|^{3/2} - \frac{ e^2}{2\epsilon_0}\sigma_0 z \sigma(z) - \mu |\sigma(z)|   
  \nn\\ &- 
\frac{ e^2}{4\epsilon_0}\int_0^D \frac{dz'}{d} \sigma(z) \sigma(z') |z - z'| \bigg\}. \label{6}
\end{align}
The function $ f(z) \equiv |\sigma(z)|^{1/2} $ which minimizes the potential $ \Omega, $ obeys the equation:
\be
f(z) - \tilde \mu - \tilde\beta\mbox{sign}[\sigma(z)] \left\{\sigma_0 z + \int_0^D \frac{dz'}{d} \sigma(z') |z - z'| \right\} = 0.  \label{7}
\ee
Here, $\mbox{sign}(x) $ is the sign function, $ \tilde \mu = 2\mu/3\gamma $ and $\tilde\beta =  e^2/3\gamma\epsilon_0. $ The rescaled chemical potential $ \tilde \mu $ has the dimensions of the inversed length whereas $ \tilde\beta $ is a dimensionless parameter. The parameter $ \tilde\beta $ measures the ratio of the Coulomb interactions strength to the kinetic energy of the charge carriers in the graphene layers. The Eq. (\ref{7}) is nonlinear with respect to $\sigma(z)$, and this reflects the essential nonlinearity of the Thomas-Fermi theory as applied to graphene packs. The corresponding solution for a conventional material (either conductor or insulator) should be linear with respect to the charge carriers density. The current nonlinearity occurs due to the particular form of the  charge carriers spectra in graphene, which manifests itself in the unusual expression for the kinetic energy given by the Eq. (\ref{1}).

Carrying out two successive differentiations  with respect to the variable  $''z''$ and using  the electroneutrality conditions given by the Eq. (\ref{5}) one may transform the integral equation (\ref{7}) to the nonlinear differential equation of the second order for the function $ f(z): $
\be
\frac{d^2 f}{dz^2} = \frac{2\tilde\beta}{d} f^2(z)  \label{8}
\ee
with the boundary conditions:
\begin{align}
&\frac{df}{dz}\Big|_{z=0} = - \tilde\beta\sigma_0 \mbox{sign} (\sigma_0),  \nn\\
&\frac{df}{dz}\Big|_{z=D} = \tilde\beta\sigma_0 \mbox{sign} (\sigma_0).  \label{9}
\end{align}
One may note that the function $f(z) $ has a negative slope at $ z=0$ and positive slope at $ z = D $ and these slopes are equal in magnitude. Also, one may expect the charge-carriers density $ \sigma(z) $ to take on equal in magnitude and opposite in sign values at the interfaces $ \big(\sigma(0) = - \sigma(D)\big). $ This gives grounds to conclude that the function $ f(z) $ reaches its minimum in the middle of the film at $ z = D/2 ,$ whereas $ \sigma (z) $ should increase/decrease over the whole interval $ 0<z<D $ depending on the sign of $ \sigma_0. $ When $ \sigma_0 > 0, \ \sigma(0)$ should take on a negative value to balance the charge on the substrate,  which means that $ \sigma(z) $ is an increasing function. On the contrary, $ \sigma(z) $ should monotonously decrease over the interval $ 0< z< D $ if $\sigma_0 < 0. $ At the middle point $ (z = D/2), \ \sigma(z) $ becomes zero, and the derivative $ df/dz $ satisfies the following relation:
\be
\frac{df}{dz}\Big|_{z=D/2 -0} = - \frac{df}{dz}\Big|_{z = D/2 +0} . \label{10}
\ee 

One cannot analytically solve the differential equation (\ref{8}) following a straightforward way. However, it could be shown (see Ref. \cite{8}) that this equation is equivalent to a conservation law of the form:
\be
\frac{d}{dz}\left\{\frac{1}{2}\left(\frac{df}{dz}\right)^2 - \frac{2\tilde\beta}{3d}f^3(z)\right\} =0.   \label{11}
\ee
Using this conservation law we obtain:
\be 
\left(\frac{df}{dz}\right)^2 = \frac{4\tilde\beta}{3d} f^3(z) + C  \label{12}
\ee
where the constant $C $ is determined by the boundary conditions. We start to analyze the solutions of the Eq. (\ref{12}) by splitting the original range $ 0\leq z \leq D $ in halves and separately solving this equation for these halves. Employing the boundary conditions (\ref{9}) and the relation (\ref{10}) and introducing a dimensionless parameter $ R $ defined by the expression:
\be
1 + R^3 = \frac{3\tilde\beta d\sigma_0^2}{4f^3(0)}  \label{13}
\ee
we may present the solution of the Eq. (\ref{12}) as follows:
\be
\frac{z}{D} =\frac{\ds \frac{1}{2} \int_r^1 \!\! \frac{du}{\sqrt{u^3 + R^3}}}{\ds \int_0^1 \! \frac{du}{\sqrt{u^3 + R^3}}},  \qquad 0\leq z \leq \frac{D}{2}  \label{14}
\ee
and
\be
\frac{z}{D} = \frac{1}{2} \left( 1 + \frac{\ds\int_0^r  \frac{du}{\sqrt{u^3 + R^3}}}{\ds \int_0^1 \! \frac{du}{\sqrt{u^3 + R^3}}}\right),  \quad  \frac{D}{2} < z \leq D. \label{15}
\ee
In these expressions, $ r(z/D) = f(z)/f(0).$ Now, it is necessary to clarify the physical meaning of the parameter $ R $ by relating it to certain characteristics describing properties of the considered system. Using Eqs. (\ref{12}),(\ref{13}) one may derive the following expression:
\be
(1 + R^3)^{1/6} \int_0^1 \frac{du}{\sqrt{u^3 + R^3}} = \Gamma  \label{16}
\ee
where
\be
\Gamma = \left(\frac{\tilde\beta^2 |\sigma_0| D^3}{6d}\right)^{1/3}. \label{17}
\ee

\begin{figure}[t] 
\begin{center}
\includegraphics[width=4.5cm,height=4.5cm,angle=-90]{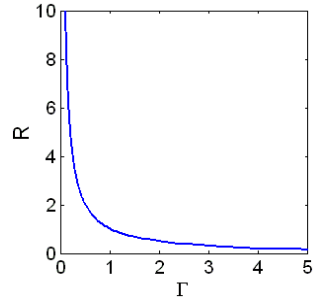}
\caption{(Color online) The parameter $ R $ as function of $ \Gamma. $ The curve is plotted using the Eq. (\ref{16}). In the strong coupling regime $(\Gamma \gg 1)\ R$ takes on values much smaller than $1$ being proportional to $\Gamma^{-2}. $ In the weak  coupling regime $(\Gamma \ll 1)\ R $  could be approximated by $1/\Gamma. $
}
 \label{rateI}
\end{center}\end{figure}

The newly introduced dimensionless parameter $ \Gamma $ is determined by the thickness of the FLG pack, the distance between the adjacent layers and by the charge carriers density on the substrates $\sigma_0.$ Also and most importantly, it depends on the parameter $\tilde \beta $ which characterizes the Coulomb interactions between the graphene layers and the substrate as well as interactions between different graphene sheets. All these characteristics are combined into the single  control parameter, which determines the nature of the FLG screening. When the system is in a strong coupling regime dominated by its electrostatic energy $ (\tilde\beta > 1),$ the parameter $ \Gamma $ may take on values significantly greater than $1. $ In the weak coupling regime, when the kinetic energy predominates $ (\tilde\beta \ll 1) $ the control parameter $ \Gamma $ should become much smaller than one. As follows from the Eq. (\ref{16}), within the extremely strong coupling limit $ (\Gamma \gg 1)$ the parameter $ R $ is inversely proportional to $ \Gamma^2 ,$ and it takes on values close to zero. In this case we may approximate $ f(0) $ as:
\be
f(0) \approx f_0 = \left(\frac{3d}{4}\tilde\beta\sigma_0^2\right)^{1/3}. \label{18}
\ee
Within the weak coupling regime $(\Gamma \ll 1) \ R \approx \Gamma^{-1}.$ Substituting this approximation into the expression for $f(0) ,$ we find: 
\be
f(0) \approx \frac{\tilde\beta|\sigma_0|D}{4}.  \label{19}
\ee
Comparing these asymptotic expressions, we see that in the weak coupling limit the ratio $ f(0)/f_0 $  takes on small values of the same order as $\Gamma. $ This means that the FLG charge carriers densities induced at the interfaces by the substrate charge carrier density $ \sigma_0 $ are significantly greater when the Coulomb interactions in the considered system are strong enough for the inequality $ \Gamma > 1 $ to be satisfied. In general case, the dependence of $ R $ on $ \Gamma $ as given by the Eq. (\ref{16}), is presented in the Fig. 1. One may observe that the crossover between the weak $(\Gamma \ll 1)$ and strong $ (\Gamma \gg 1)$ coupling regimes occurs at $ R \sim 1. $

 The charge carriers density in the FLG is simply related to the function $ r(z/d).$ Assuming for certainty that $ \sigma_0 > 0 $ we obtain:
\be 
\sigma\left(\frac{z}{D}\right) = \left\{  \ba{l}\ds
- \frac{f_0^2 }{\big(1 + R^3\big)^{2/3}}\, r^2 \left(\frac{z}{D}\right),  \qquad 0 \leq z \leq \frac{D}{2},  \\ \ds \ \
\frac{f_0^2 }{\big(1 + R^3\big)^{2/3}}\, r^2 \left(\frac{z}{D}\right),  \qquad \frac{D}{2} \leq z \leq D.
\ea   \right.  \label{20}
\ee
The functions $ r(z/D) $ and $\tilde\sigma(z) = \sigma(z/D)\big/\sigma(0) $ determined  by the Eqs. (\ref{14}),(\ref{15}) and (\ref{20}) are plotted in the Fig. 2 for several values of the parameter $ R. $ One may observe that within the weak coupling regime the magnitude of $ \sigma(z) $ rather slowly changes as one moves away from the interfaces into the FLG pack interior. This indicates that the screening length of the external charge on the substrates is long, which can be expected since graphene sheets are semimetallic. Actually, $ r(z/D) $ is almost independent of $ R $ if this parameter takes on values greater than one, and it could be approximated as follows:
\be 
r\left(\frac{z}{D}\right) = \left\{  \ba{l}\ds
1 - \frac{2z}{D}  \qquad \left(0 \leq z \leq \frac{D}{2}\right),  
\\ \ds
\frac{2z}{D} - 1  \qquad \left(\frac{D}{2} \leq z \leq D\right) 
\ea   \right.  \label{21}
\ee
which results in the square-law dependence of $ \tilde \sigma $ on $ z/D.$

\begin{figure}[t] 
\begin{center}
\includegraphics[width=4.7cm,height=9.3cm,angle=-90]{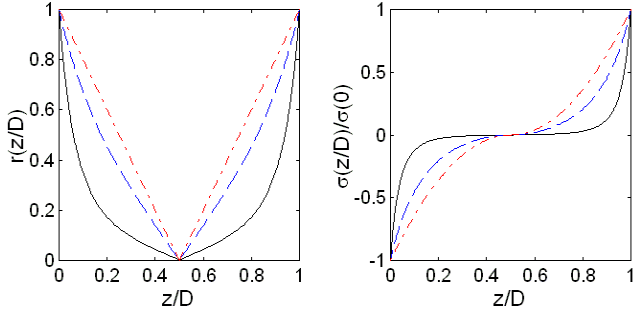}
\caption{(Color online) Ratio $ f(z)/f(0) $ (left panel) and $\sigma(z)/\sigma(0) $ (right panel) as a function of the normalized distance into the film. The curves are plotted in accordance with Eqs. (\ref{14}),(\ref{15}) and (\ref{20}) assuming $R = 10 $ (dash-dotted lines), $R = 1 $ (dashed lines) and $ R = 0. 1$ (solid lines) which corresponds to the weak, intermediate and strong coupling regimes, respectively.
}
 \label{rateI}
\end{center}\end{figure}

On the contrary, within the strong coupling regime the FLG charge carriers density magnitude exhibits a steep decrease as we move into the film. In this case the major portion of the induced charge is concentrated in the close vicinities of the interfaces whereas the FLG interior remains nearly neutral. So,  the FLG behaves as a conductor where the external charge is efficiently screened by a surface charge distribution confined to the bounded graphene layer. Such metallic-like behavior may occur when $ \Gamma \gg 1, $ which could be provided by the sufficiently large amount of charge put into the substrates (large $ \sigma_0)$ on condition that the graphene layers in the FLG are densely packed, so the electrostatic potential energy predominates over the kinetic term $(\tilde\beta > 1). $

The obtained results enable us to  analyze the electrostatic potential profile across the FLG film. This occurs because the renormalized electrostatic potential $ \tilde\Phi (z) = 2\Phi(z)/3\gamma $  given by the expression:
\be
e\tilde\Phi(z) = -\tilde\beta z \sigma_0  \mbox{sign} [\sigma(z)] - \tilde\beta\mbox{sign}[\sigma(z)] \int_0^D \frac{dz'}{d}\sigma(z') |z - z'| \label{22}
\ee
is simply related  to the function $ f(z). $ Comparing this expression and the Eq. (\ref{7}) we get:
\be
e\tilde\Phi (z) = \tilde\mu - f(z).  \label{23}
\ee
Assuming for certainty that $ \Phi(0) = 0 $ we easily find the corresponding value of the chemical potential $\tilde\mu. $ Substituting the result into  Eq. (\ref{23}) we obtain the following expression for the dimensionless quantity $e\tilde\Phi(z)/f_0 $ which is closely related to the electrostatic potential:
\be
\frac{e\tilde\Phi(z)}{f_0} = \frac{\ds 1 + \mbox{sign}\left[z - \frac{D}{2}\right] r\left(\frac{z}{d}\right)}{\big(1 + R^3\big)^{1/3}} . \label{24}
\ee
The profile of the electrostatic potential  strongly depends on the coupling regime. When the electrostatic interactions in the system are weak $(\Gamma \ll 1),$ we may employ the approximations (\ref{21}) for the function $r(z/D).$  In this case the electrostatic potential has a linear profile and the potential difference across the film equals: 
\be
\Phi(D) - \Phi(0) \equiv \Delta\Phi =  \frac{e\sigma_0 D}{4\epsilon_0}.  \label{25}
\ee
So, when the Coulomb interactions are weak, the considered system behaves as a parallel-plate capacitor, and the FLG takes on the part of a dielectric material filling the space between the plates and characterized by the dielectric constant $ \kappa = 2. $

However, when the parameter $\Gamma $ increases the approximation given by the Eq. (\ref{25}) ceases to be valid. In general case the electrostatic potential difference is given by the expression
\be
\Phi(D) - \Phi(0) = \frac{3\gamma}{e}\frac{f_0}{(1 + R^3)^{1/3}}  \label{26}
\ee
and it exhibits a nonlinear dependence on $ \sigma_0. $ Within the strong coupling limit $(R\ll 1) $ the potential difference between the interfaces $ \Delta\Phi $ is proportional to $ \sigma_0^{2/3},$ so one may consider the system as a capacitor whose differential capacitance $ C $ varies as $ \Delta \Phi $ changes, being proportional to $(\Delta \Phi)^{1/2}.$ The  electrostatic potential profiles are presented in the left panel of the  Fig. 3. In a weak coupling the potential increases nearly linearly as we move into the FLG film. The stronger is the coupling the more pronounced is the potential change in the vicinities of the interfaces. One may expect that in the limit of the very strong coupling $(R \ll 1)$ almost the whole potential drop should occur near the interfaces leaving the potential nearly constant in the main body of the film.  

\begin{figure}[t] 
\begin{center}
\includegraphics[width=4.7cm,height=9.3cm,angle=-90]{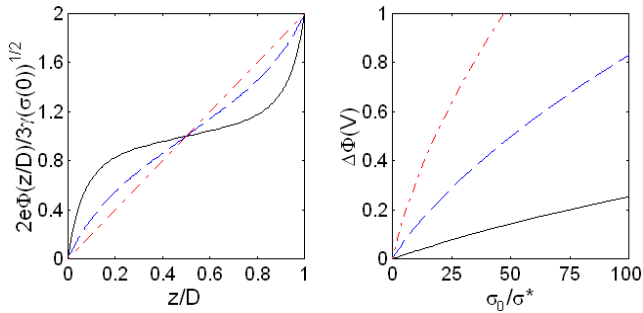}
\caption{(Color online) Left panel: The spatial profiles of the scaled electrostatic potential inside the FLG film as functions of the normalized distance into the film plotted using Eq. (\ref{24}) for $ R = 10$ (dash-dotted lines), $R = 1 $ (dashed lines) and $ R = 0. 1$ (solid lines). Right panel: The electrostatic potential difference $ \Delta \Phi $ as a function of the normalized areal charge density on the substrates $ \sigma_0 .$ The curves are plotted in accordance with Eqs. (\ref{16})-(\ref{18}), and (\ref{25}) assuming that $\gamma = 6.5 eV,\ d = 3.347 \mathring A,\ \sigma^* = 10^{-4} 1/\mathring A^2 $ for $D = 7d $ (dash-dotted line), $ D = 5 d$ (dashed line), and $ D = 3d $ (solid line).
}
 \label{rateI}
\end{center}\end{figure}

One may increase the parameter $ \Gamma $ and thus stimulate the switching to the regime of strong coupling following two ways. First, one can enhance $ \sigma_0 $ keeping $ D $ fixed. This would be an appropriate analysis for experiments on a single FLG sample of a certain thickness where $\sigma_0 $ is varied by varying the voltage applied across the system. Secondly, one may keep $\sigma_0 $ fixed and increase the film thickness $ D $ by adding extra graphene sheets to the pack. We illustrate the effect of these two factors on the electrostatic potential change across the film in the right panel of the Fig. 3. Curves shown in this figure correspond to different values of $ D. $ At small values of $\sigma_0 $ all curves exhibit nearly linear dependencies of $\Delta\Phi $ on $\sigma_0 $ which is typical for the weak coupling regime. As $\sigma _0 $ increases the curves deviate from the corresponding straight lines, and this indicates the transition to the strong coupling regime. We observe that the greater becomes $ D $ the smaller $ \sigma_0 $ is required to provide this transition. We remark that the slopes of the presented curves at the $\sigma_0 = 0 $ differ for different values of $D. $ The greatest value of $ D $ corresponds to the steepest slope, which obviously makes sense. As follows from Eq. (\ref{26}) in the strong coupling regime $(\Gamma \gg 1) $ the electrostatic potential difference is close to $\frac{3\gamma}{e}f_0, $ and the first correction to this main approximation is proportional to $ D^{-6}. $

To further analyze the possibilities for experimental observations of the described features in the electric charge and electrostatic potential distribution in the FLG samples, one should take into account the direct contribution to the potential from the substrates. For certainty, we assume that the areal charge  densities on the surfaces of the substrates appear due to the presence of ionized impurities in the substrate materials. The impurities  are supposed to be uniformly distributed in the substrates with the volume density $\rho_0. $ First, we consider the substrate layer adjoining the FLG at $z = 0. $ Then the electrostatic potential inside the substrate layer obeys the Poisson equation which in the considered case is reduced to the form:
\be
\frac{d^2\Phi_s}{dz^2} = - \frac{e}{\epsilon_0}\rho_0.  \label{27}
\ee
Setting $\Phi_s = 0 $ at the depletion length $L_D $ we obtain the solution of this equation, namely:
\be
\Phi_s = - \frac{e\rho_0}{2\epsilon_0}(z + L_D)^2.  \label{28}
\ee
Calculating the electric field at the interface $z = 0\ (E = -\partial\Phi_s/dz) $ and keeping in mind that $E(0) = e\sigma_0/\epsilon_0 $ we find that $ L_D = \sigma_0/\rho_0. $ Accordingly, the electrostatic potential $\Phi_s $ at the interface equals:
\be
\Phi_s(0) = - \frac{e \sigma_0^2}{2\epsilon_0\rho_0}.  \label{29}
\ee
The electrostatic potential associated with the substrate at the substrate/FLG interface at $ z= D $ could be estimated following a similar way. As a result, the total electrostatic potential difference including the contribution from the FLG as well as from the substrate takes on the form:
\be
\Delta\Phi = \frac{3\gamma}{e}\frac{f_0}{(1+R^3)^{1/3}} + \frac{e\sigma_0^2}{\epsilon_0\rho_0}.  \label{30}
\ee
Within the  low coupling limit the ratio of the terms  in the Eq. (\ref{30}) is of the order of $ D\rho_0/\sigma_0 $ (see Eq. (\ref{25})), therefore the first term predominates if $ \rho_0 > \sigma_0/D. $ If the Coulomb interactions are strong the FLG contribution dominates when
\be
\rho_0 > \left(\frac{4}{3d}\tilde\beta^2 \sigma_0^4 \right)^{1/3}.  \label{31}
\ee
The expression for the total potential drop (\ref{30}) must be changed if we allow for the predominating effect of surface/interface states. This seems  a likely resolution with the culprit being some adsorbed species confined to the interfacial layer. This basically cuts off the depletion length $L_D $ at the characteristic depth for the surface states $d_s\ (d_s < L_D). $ Then $ \sigma_0 = \rho_0 d_s $ and the contribution from the substrates into the total potential drop accepts the form $ e\sigma_0 d_s/\epsilon_0. $ In this case, the predomination of the FLG contribution to the total potential drop becomes easier to reach. In weak coupling regime the graphene contribution  predominates when $ d_s < D $ which is usually the case. When the coupling is strong, one must require the inequality
\be
d_s <\left(\frac{3d}{4\tilde\beta^2 \sigma_0}\right)^{1/3}   \label{32}
\ee
to be satisfied to provide the prevalence of the graphene contribution to the potential drop.
\vspace{3mm}

 The results of the present work could be summarized as follows. Since a single graphene sheet is semimetallic one may expect the screening length in a FLG film placed between two charged substrates (or exposed to an external electric field in some other way) to be long. In consequence, any electric charge donated to the film would be distributed over all graphene layers, and the electrostatic potential inside the film would have a linear spatial profile. However, these conclusions remain valid only within the weak coupling regime $ (\Gamma \ll 1), $ when the Coulomb interactions between the graphene sheets as well as between the film and the substrates are smaller than the kinetic energy of the charge carriers. Increasing the areal carriers density on the substrates and/or making the FLG sample more densely packed be reducing the ratio $d/D, $ one may carry out a transition to the strong coupling regime $(\Gamma > 1) .$  Within this regime, the FLG exhibit a metallic-like behavior, so that the induced electric charge becomes concentrated near the surfaces of the FLG film, and the electrostatic potential inside the film acquires a nonlinear spatial profile. The transition from weak to strong coupling regime turns on relatively quickly at a crossover value of the external areal charge density $ \sigma_0 $ (or at a crossover value of the external electric field).  The obtained results ignore the effects of hybridization between the adjacent graphene sheets. Therefore, these results could be applied either to loosely packed FLG samples (where the interlayer tunnelings are prevented by the sufficient layers separation) or to twisted graphenes where the interlayer tunneling are hindered due to the rotational misorientation of the crystalline lattices.

Finally, we estimate the parameters characterizing the considered system to show that the strong coupling regime where the graphene sheets exhibit a metallic-like behavior could be reached in realistic experiments. The typical value of the parameter $ \gamma $ is $ 6.5 eV  \mathring A $ which gives for the dimensionless constant $\tilde\beta $ the value of the order of ten, namely: $ \tilde\beta =9.5. $ Then assuming that $ d$ takes on a value close to the interlayer spacing in graphite $(d \approx 3.347 \mathring A)$ the crossover value of the parameter $ \Gamma\ (\Gamma = 1) $ occurs at $0.5 \sigma_0 \approx 3d/2\tilde\beta^2 D^3 \approx 6.5\times 10^{-2}/D^3. $ This areal  density has dimensions $ \mathring A^{-2} $ if the FLG thickness is expressed in $ \mathring A. $ The minimum value of $ D $ cannot be smaller than $ d, $ so $\frac{1}{2} \sigma_0 $ needs only to exceed $ \sim 2\times 10^{-3} \mathring A^{-2} $ to provide the switching of the system to the strong coupling regime. We remark that for $ D > d $ this crossover value for $ \sigma_0 $ may be one or even two orders of magnitude smaller. Certainly, such values of the areal charge carriers densities at the surfaces of substrates are within the experimentally accessible range. However, the specific features originating from the special properties of graphene could be manifested only provided that the contributions from the substrates to the total  electrostatic potential distribution in the system are small compared to the contribution from the FLG. As follows from the Eqs. (\ref{30}),(\ref{31}) this occurs when the volume density of charged impurities uniformly distributed in the substrates $ \rho_0 \sim 10^{20} 1/cm^3 $ (assuming that $ \frac{1}{2}\sigma_0 \sim 10^{-3} \mathring A^{-2}),$ which is quite large. One may obtain smaller and more realistic values for $ \rho_0 $ by reducing $ \sigma_0 .$ However, following this way  one may find oneself beyond the strong coupling regime which is the most interesting for observation. The situation is much better in the case when the areal charge densities at the substrate surfaces appear due to the effect of surface states with the characteristic depth $ d_s. $ Using Eq. (\ref{32}), we may estimate $d_s \sim 1nm $ at $ \frac{1}{2}\sigma_0 \sim 10^{-3} \mathring A^{-2}. $ This is quite reasonable. The substrate only gets to dominate if the characteristic depth $d_s $ becomes significantly longer than one estimated above. So, the specific effects originating from the particular charge carriers spectra in graphene which were discussed in the present work are likely to be accessible for experimental observations.

{\it Acknowledgments:} 
The author thanks E. J. Mele for helpful discussions and  G. M. Zimbovsky for help with the 
manuscript. The work was partly supported  by  NSF-DMR-PREM 0353730.



\begin{thebibliography}{99}

\bibitem{1} K. S. Novoselov et al, Science {\bf 306}, 666 (2004).

\bibitem{2} J. R. Williams, L. DiCarbo, and C. M. Marcus, Science {\bf 317}, 638 (2007).
 
\bibitem{3} Z.Luo, L. A. Somers, Y. Dau, T. Ly, N. J. Kybert, E. J. Mele, and A. T. Johnson, Nano Lett. {\bf 10}, 777 (2010).

\bibitem{4} F. Guinea, Phys. Rev. B {\bf 75}, 235433 (2007).
 
\bibitem{5} H. Min, B. Sahu, S. K. Banerjee, and A. H. MacDonald, Phys. Rev. B {\bf 75}, 155115 (2007). 

\bibitem{6} S. Y. Zhou et al, Nat. Matter {\bf 6}, 770 (2007).
 
\bibitem{7} T. Ohta et al, Phys. Rev. Lett. {\bf 98}, 206802 (2007).
 
\bibitem{8} S. S. Datta, D. R. Strachan, E. J. Mele, and A. T. Johnson, Nano Lett. {\bf 9}, 7 (2009).

\bibitem{9} W. A. de Heer, C. Berger, X. Wu, M. Sprinkle, Y. Hu, M. Ruan, J. A. Stroscio, P. N. First, R. Haddon, B. Piot, C. Faugeras, M. Potemski, J-S. Moon, 	J. Phys. D: Appl. Phys. {\bf 43}, 374007 (2010). 

\bibitem{10} J. Hass, F. Varchon, J. E. Millán-Otoya, M. Sprinkle, N. Sharma, W. A. de Heer, C. Berger, P. N. First, L. Magaud, and E. H. Conrad, Phys. Rev. Lett. {\bf 100}, 125504 (2008).

\bibitem{11} A. Reina, X. Jia, J. Ho, D. Nezich, H. Son, V.Bulovic, M. S. Dresselhaus, and J. Kong, Nano Lett.{\bf 9}, 30 (2009).

\bibitem{12} H. Schmidt, T. Luedtke, P. Barthold, E. McCann, V. I. Falko, R. J. Haug, Appl. Phys. Lett. {\bf 93}, 172108 (2008).

\bibitem{13} G. Li, A. Luican, and E. Y. Andrei, Phys. Rev. Lett. {\bf 102}, 176804 (2009).

\bibitem{14} L. Pietronero, S. Strassler, H. R. Zeller, and M. J. Rice, Phys. Rev. Lett {\bf 41}, 763 (1978).

\bibitem{15} S. A. Safran and D. R. Hamann, Phys. Rev. B {\bf 22} 606 (1980).



\end{thebibliography}
\end{document}